# Understanding variability in HASDM to support space traffic management

**W. Kent Tobiska, Marcin D. Pilinski, Shaylah Mutschler,
Kaiya Wahl, Jean Yoshii, Dave Bouwer**
*Space Environment Technologies*

**Piyush Mehta and Richard Licata**
*West Virginia University*

## ABSTRACT SUMMARY

Today we see an expanding drive for global, high-speed internet, which is the de-facto life blood of the global economy and of our national defense. The demand for higher internet speeds with global internet connectivity, anywhere, anytime, touches every part of our technological society. Commercial efforts, such as SpaceX's Starlink constellation, but not limited to them, are now hugely expanding global internet capacity in the next two years. With more constellations planned, the number of Low Earth Orbit (LEO) objects is set to TRIPLE in two years. The growth in the number of LEO objects directly increases the probability of unintentional collisions between objects due to accumulating space debris. This is the runaway Kessler syndrome, where more and more unavoidable collisions occur and leads to a potentially unusable LEO orbital domain. Effective space traffic management needs accurate knowledge of the variability in upper atmosphere densities. Data assimilative modeling, where physics-based models are informed by measurements, supplies the best capability today for specifying and predicting space weather. Ensemble modeling, where multiple models are run for an epoch, is an excellent method for characterizing the range of variability. The foundation for this modeling comes from the SET High Accuracy Satellite Drag Model (HASDM) density database. We report on studies to understand the variabilities in HASDM, and this knowledge support operational space traffic management and collision assessment activities. We identify two features about the thermosphere from the SET HASDM density database. First, we have confirmed that the time scale is very rapid (1-h) for molecular conduction above 200 km to transfer energy vertically in the thermosphere. This results couples with a longer timescale for conduction in the 100–200 km region where it takes up to 2 days for energy to transition across that region via molecular conduction. We now have, for the first time, an excellent picture of the timescales of energy change (dE/dt) throughout the thermosphere. Second, the SET HASDM density data display a common range of variability despite the level of daily averaged geomagnetic activity as represented by Ap. During higher levels of daily averaged geomagnetic activity, the density mean and median values increase at all altitude levels. However, the relative range of variability is consistent from one daily average of Ap to the next. The reason for this is likely to be that the underlying pre-storm density of the thermosphere is determined by the solar EUV and FUV irradiances that create a thermal foundation of the upper atmosphere. The daily averaged geomagnetic activity is a perturbation upon that foundation. The global range of variation occurs because geomagnetic activity separately modifies the underlying thermal inertia in the atmosphere. The mean and median density increases are related to geomagnetic storms.

## 1. INTRODUCTION

**Background.** Today we see an expanding drive for global, high-speed internet, which is the de-facto life blood of the global economy and of our national defense. The demand for higher internet speeds with global internet connectivity, anywhere, anytime, touches every part of our technological society. Commercial efforts, such as SpaceX's Starlink constellation, but not limited to them, are now hugely expanding global internet capacity in the next two years. With more constellations planned, the number of Low Earth Orbit (LEO) objects is set to TRIPLE in two years. The growth in the number of LEO objects directly increases the probability of unintentional collisions between objects due to accumulating space debris. The growth in the number of LEO objects directly increases the probability of unintentional collisions between objects due to accumulating space debris. This is the runaway Kessler syndrome, where more and more unavoidable collisions occur, leading to a potentially unusable LEO orbital domain.

The NASA LEGEND orbital debris program was run in the early 2000's and it estimated the growth in collision fragments if there were no launches after 2000 (Figure 1, red curve) [https://www.quora.com/What-is-the-likelihood-of-Kessler-syndrome-happening-around-the-Earth-anytime-in-the-near-future]. The projected rise of collision fragments continues for the next 200 years in that model run. To state the obvious, we have not zeroed out launch rates since 2000 but have increased them. The actual number of LEO objects today are a factor of two higher than those shown in Figure 1. This figure ironically shows the start of collision fragment rise in 2009, which is when the unintentional collision between Iridium 33 and KOSMOS 2251 occurred (February 10, 2009). Since then, the debris environment has worsened with Chinese, Indian, and Russian ASAT tests. The analysis also does not include thermospheric cooling, whereby the lower thermosphere cools due to $CO_2$ accumulation from below and leads to a cooler thermosphere with lower densities, resulting in longer debris lifetimes. In Figure 2, we see the debris size vs. population for LEO objects. The dark red circle outlines a particularly dangerous population that is 1–10 cm in diameter; it is barely tracked but capable of significant spacecraft subsystem damage or even catastrophic loss for active vehicles.

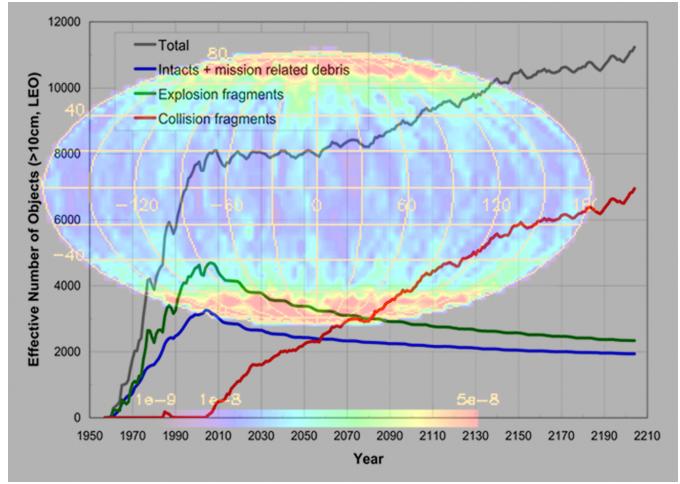

Fig. 1. Projected Kessler syndrome from NASA's LEGEND model.

**Debris risk reduction.** From the *National Orbital Debris Research and Development Plan* (NODRDP, 2021) [1] there are three ways to reduce the risks associated with debris:

1. **Limit the generation of new debris.** Referencing Table 1 in the NODRDP, items 1.1–1.6 for limiting by design the generation of new debris includes: *i)* reduced debris during launch, *ii)* improved resilience of spacecraft surfaces; *iii)* improved shielding and impact resistance; *iv)* developed designs that will reduce or limit fragmentation process; *v)* improved maneuverability capabilities; and *vi)* incorporated end-of-mission approaches to minimize debris into spacecraft and mission design. The ISO standard *ISO 24113: Space Systems – Space debris mitigation requirements* [2] details these strategies and best practices. As an example, the Starlink, Kuiper, and OneWeb constellations are joining a satellite and debris domain that will experience active oversight within the emerging Space Traffic Management (STM) paradigm of the Department of Commerce (DoC). Hundreds of these satellites are managed with operational procedures to deorbit when the useful lifetime has completed.

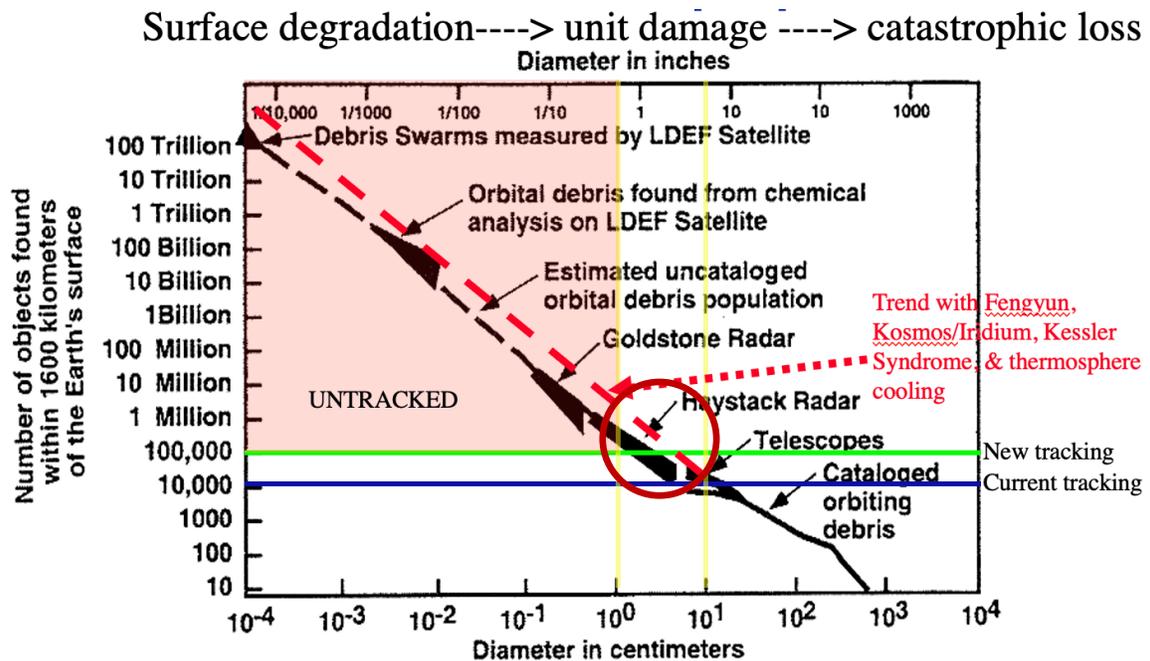

Fig. 2. Debris size vs. population count for objects in LEO.

2. **Better track and characterize debris.** It is important to know the percent error of atmospheric density, which is a key item needed to account for mismodeling of the current epoch and forecast atmosphere density [3]. If the atmosphere density uncertainty is known, then the square of that uncertainty can is added to the normalized ballistic coefficient variance. The result is a probability of collision between two satellites that is more accurately predicted. If this mismodeling can be corrected by the provision of density uncertainties, operators can then better assess the uncertainties in the thermosphere and improve their Conjunction Assessment (CA) strategies.

For the foundation of atmospheric density, all STM systems, including NASA's Conjunction Assessment Risk Analysis (CARA) program supporting CA, depend upon the functioning of one thing – the U.S. Space Force (USSF) Space Command's High Accuracy Satellite Drag Model (HASDM) [4, 5]. HASDM is a data assimilative system combined with an empirical forecast model. It supplies the operational world's most accurate atmospheric density to its users. However, it is only as good as its underlying uncertainty. Space Environment Technologies' (SET) META-HASDM project [6] is developing HASDM driver updates with uncertainties that are quantified. This activity addresses the need for a capability to better track and characterize debris (see NODRDP Table 1 items 2.1–2.3: *i)* characterizing orbital debris and the space environment; *ii)* developing technologies to improve orbital debris tracking and characterization; and *iii)* reducing uncertainties of debris data in orbit propagation and prediction). For the first time, META-HASDM is providing the uncertainties associated with atmosphere densities in different altitude layers as well as the forecast solar and geomagnetic indices drivers. These uncertainties, discussed below, are being included in a metadata section of the operational files delivered by SET to USSF. The net result will be a reduction of the error ellipsoids in the conjunction assessment process for LEO objects.

3. **Remediate existing debris.** Because of the increasing LEO object population, especially those of 10 cm or less in diameter, their removal is an important goal since they can cause catastrophic damage to active orbital assets. Larger objects such as rocket bodies and dead satellites can be remediated with rendezvous techniques. However, the removal of the smaller object population is just as important since these objects number in the hundreds of thousands to over a million in LEO. Unique technologies to remediate this smaller object population are needed, including methods such as beamed photon radiation pressure in a direction that opposes the orbital velocity to contribute an additive force with atmospheric drag to dissipate orbital energy. The result, through time, leads to the gradual removal of small LEO objects sooner than would occur naturally. This method takes advantage of atmospheric drag's natural effects and nudges small objects into lower, more dense atmosphere layers. Figure 3 shows that solar radiation pressure (SRP) at 500 km during solar minimum with a 700 K exospheric temperature has the same force as atmospheric drag. At 650 km the SRP has the same force as during solar maximum conditions with a 1000 K exospheric temperature. The dark red oval shows the forces' domain.

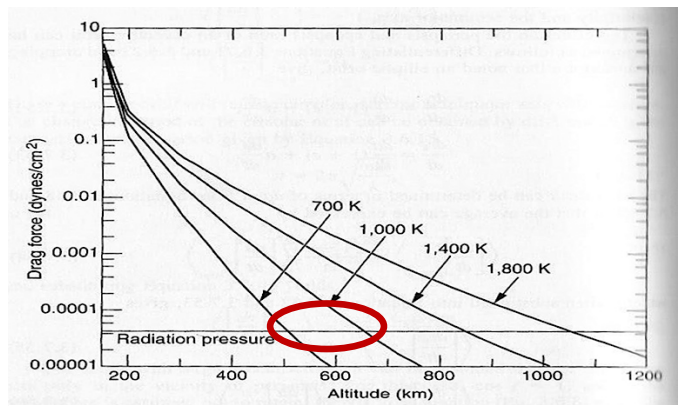

Fig. 3. Atmospheric drag vs. solar radiation pressure in LEO for solar minimum (700 K) to solar maximum (1000 K) exospheric temperatures. Credit Piscane.

## 2. METHODOLOGY

**HASDM uncertainties.** To understand the atmosphere density uncertainties, these questions are relevant: "*What are the absolute uncertainties in HASDM relevant for understanding collision avoidance*" and "*what ar the ranges of variability in the thermosphere under various conditions?*" The SET HASDM density database [5] is unique in that it supports the most accurate representation of the global thermosphere densities for the 20-year period from January 1, 2000, to December 31, 2019, with 3-h time resolution, 25-km altitude steps and a 175–825 km altitude range on a 10° × 15° latitude/longitude grid. This database is publicly available for scientific research use and is found at the URL https://spacewx.com/hasdm/.

The HASDM data assimilation system changes the JB2008 model densities using a dynamic calibration of the atmosphere (DCA) with a segmented solution approach. This approach extracts the time resolution needed to accurately determine the dynamically changing thermospheric density by taking a 3-h sub-interval within the fit span of an estimated 1.5-day interval for each of up to 90 calibration satellites. Densities are given every 3 hours. The information content in the database inherently includes all the geomagnetic storm and sub-storm, extended solar flare, nitric oxide (NO) and carbon dioxide ($CO_2$) thermospheric cooling perturbations. Because of its accuracy, time resolution, global scale, and information content, the SET HASDM database densities are now used as a space weather benchmark for atmospheric expansion. This database is refining the Phase 1 Benchmark that was released by the National Science and Technology Council [7] for upper atmospheric expansion.

Earlier work has looked at uncertainties in the HASDM densities to answer the question, *"What are the absolute uncertainties in HASDM relevant for understanding collision avoidance."* For example, the statistical uncertainty of the HASDM density database was found from machine learned analysis [8, 9] and the range of error in the HASDM database was quantified using comparisons of HASDM annual densities to the percentage of density error in well-known calibration satellites [5]. We expand upon that work here by showing two additional analyses: *i)* the range of variability in the SET HASDM density database for six unique geomagnetic storms and *ii)* the range of variability for six daily average levels of Ap index.

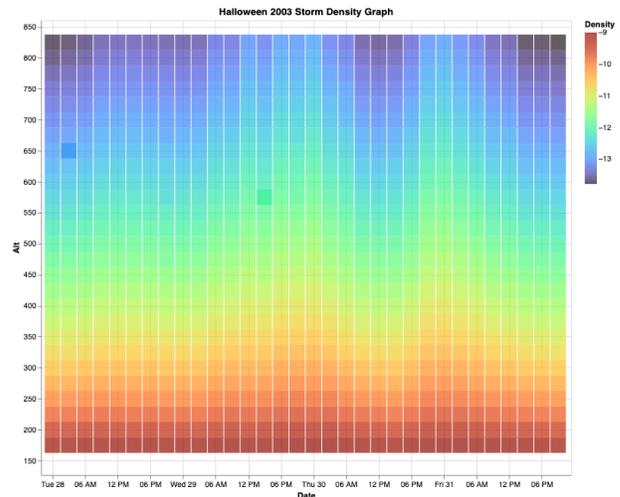

Fig 4. Halloween 2003 storm 1-σ percentage density uncertainty before, during, and after storm in No. hemisphere.

## 3. RESULTS

**HASDM storm analysis.** An analysis of HASDM density uncertainties was conducted for six storms listed in Table 1. Each storm (peak and +/- 24 hours on either side) has unique characteristics as described in the Table 1 comments. For example, the Halloween storm of October 30, 2003, is seen in Figure 4. The storm $\log_{10}$ densities are shown for 3-h time periods on the x-axis and for altitude on the y-axis. Figures 5 and 6 show the northern and southern hemisphere 1σ% density uncertainty before, during, and after the storm. The lowest uncertainty is at the lowest altitudes since the density is greatest in that region. The highest uncertainty area is the 800 km level where the densities are quiet low. As the storm proceeds, all levels of the atmosphere change and the percent uncertainty increases with altitude.

**Table 1. HASDM storm periods**

| Onset date | Min Dst (nT) | Max Kp | Max Ap | NOAA G | Peak X-ray | % variability | Comments |
|---|---|---|---|---|---|---|---|
| 2003-10-30 | -383 | 9 | 400 | 5 | X10 | +22 to +178 | Only extreme event after 1989 |
| 2003-11-20 | -422 | 9- | 300 | 4 | M4.2 | +11 to +56 | Last severe event of cycle 23 |
| 2015-03-17 | -222 | 8- | 179 | 4 | M1.8 | +7 to -10 | Shock-led multi-structure CME |
| 2015-06-22 | -204 | 8+ | 236 | 4 | M6.5 | -15 to -58 | Multiple CME impacts |
| 2015-10-07 | -124 | 7+ | 154 | 3 | NA | -9 to -74 | CIR-driven storm |
| 2017-09-08 | -122 | 8+ | 236 | 4 | X9.3 | -13 to -76 | Largest eruption of cycle 24 |

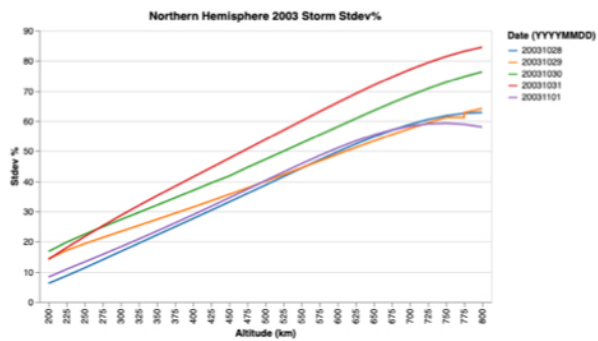

Fig 5. Halloween 2003 storm 1-σ% density uncertainty before, during, and after storm in No. hemisphere.

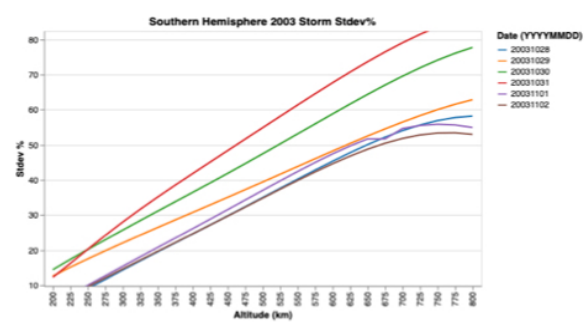

Fig 6. Halloween 2003 storm 1-σ% density uncertainty before, during, and after storm in So. hemisphere.

Figure 7 shows the peak density by altitude for the storm evolution, where there is less than an hour difference between the peak density at 200 km and the peak density at 800 km (red dashed vertical line). Thus, the time scale is very rapid for molecular conduction that transfers energy and is responsible for density surges vertically in the thermosphere. This figure confirms theoretical results for rapid molecular conduction vertically in the thermosphere [10].

In the 100–200 km in the thermosphere (not shown here) there are two lines of evidence that this region takes up to 2 days for an energy pulse to transition through the region via molecular conduction. First, JB2008 development [11] showed that an approximate 2-day lag was needed for energy to be felt by satellites above 200 km when perturbed by energy input at 100 km. The energy pulse at 100 km comes from the dissociation of $O_2$ by solar photospheric Schumann-Runge Continuum photons near the far ultraviolet (FUV) wavelengths of 160 nm. This energy is represented by the M10 solar index [12]. Second, recent work [13] shows that the SET HASDM density database responds with a negative heating (cooling) for lagged Ap values during severe geomagnetic storms. This is the phenomena where nitric oxide (NO) is produced around 100 km altitude in higher latitudes because of 300 keV electron precipitation during storms. The NO then radiates at 5.3 microns where the upwardly directed photons are lost to space and are a net cooling of the 100 km layer. This creates a 'cold plate' layer which draws down the heat from above, again by molecular conduction, and takes 36–57 hours. This is the same approximate time scale for heating due to 160 nm photons going the other direction in this region. The ~2-day molecular conduction below 200 km combines with the ~1-hour molecular conduction above 200 km and provides us, for the first time, an excellent picture of timescales for the change in energy through time (dE/dt) throughout the thermosphere.

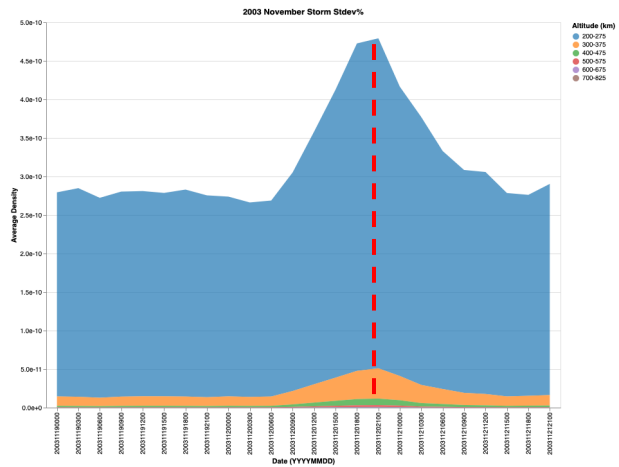

Fig 7. Halloween 2003 storm time evolution of density from 200 to 800 km – molecular conduction.

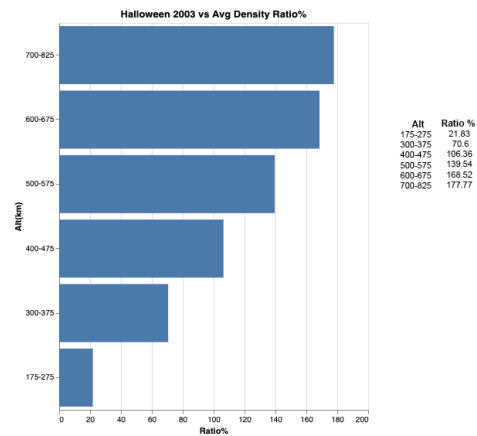

Fig 8. Halloween 2003 storm density ratios (percentage) to average storm values by altitude.

Figures 8 shows the ratio of the Halloween 2003 storm density changes compared to the average of all storms in Table 1 while Figure 9 shows the altitude scale of a storm's densities compared to the average of all storms. In Figures 10a-f we show the comparison of all six storms densities through time by altitude, in Figures 11a-f we show the individual storm ratios to the average, and in Figures 12a-f we show the six storms' densities by altitude (gold) vs. the average of all storms (black).

**HASDM daily geomagnetic condition analysis.** The analysis of the HASDM neutral density range of variability is extended to six representative daily averaged geomagnetic conditions as represented by Ap: *i)* Ap >140 (severe storm); *ii)* Ap = 85 (moderate storm); *iii)* Ap = 50 (small storm); *iv)* Ap = 27 (disturbed); *v)* Ap = 7 (quiet); and *vi)* Ap = 0 (baseline) conditions. In Figures 13a-f we show these variations. The blue bands are the range of daily, global variation for the averaged Ap value, the red line is the mean of the data, and the black line is the median of the data.

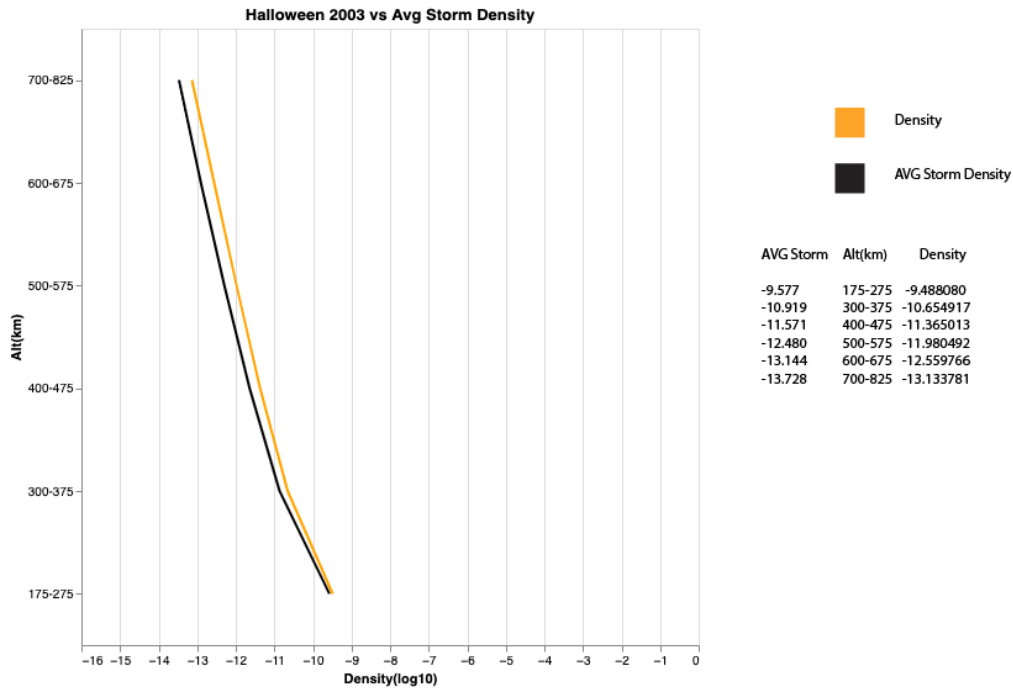

Fig. 9. Halloween 2003 storm density ratios (gold) to average storm values (black) by altitude.

In Figures 13a-f there is an interesting phenomenon that becomes apparent. The data display a common range of variability despite the level of daily averaged geomagnetic activity. As one expects, during higher levels of daily averaged geomagnetic activity the mean and median values of the density increase at all altitude levels. However, the relative range of variability tends to be consistent from one daily average of Ap to the next. The reason for this is likely to be that the underlying pre-storm density of the thermosphere, which is determined by the solar EUV and FUV irradiances, can be considered as the thermal foundation of the upper atmosphere. Thus, the daily averaged geomagnetic activity is a perturbation upon that foundation. The global range of variation shown here occurs because of the geomagnetic activity perturbations but not the underlying thermal inertia in the atmosphere. The mean and median increases in the density are geomagnetic storm related.

## 4. CONCLUSIONS

We have seen two interesting features about the thermosphere from the SET HASDM density database. First, we have confirmed that the time scale is very rapid for molecular conduction above 200 km to transfer energy vertically in the thermosphere. It takes about an hour for the increase in density, and thus, temperature, to change across 600 km of altitude during a geomagnetic storm. This confirms theoretical results for rapid molecular conduction (dE/dt) vertically in the thermosphere. These results couple with a separately identified longer timescale for conduction in the 100–200 km region where two lines of evidence show it takes up to 2 days for energy to transition across that region via molecular conduction. One line of evidence is the JB2008 time constant for the M10 solar index, which uses an approximate 2-day lag for energy transfer from 100 to 200 km. The M10 energy pulse comes from the photospheric photons from the solar Schumann-Runge Continuum at FUV wavelengths of 160 nm, which dissociate $O_2$ at 100 km. The second line of evidence is from work showing that the SET HASDM density database responds with a cooling for lagged Ap values during severe geomagnetic storms. This is the phenomena where NO is produced around 100 km altitude in higher latitudes because of 300 keV electron precipitation during storms. The NO then radiates at 5.3 microns where the upwardly directed photons are lost to space and represent a net cooling of the 100 km layer. This 'cold plate' then draws down the heat from above via molecular conduction and takes 36–57 hours. The ~2-day molecular conduction from 100–200 km is combined with the ~1-hour molecular conduction above 200 km and provides, for the first time, an excellent picture of the energy change timescales (dE/dt) in the thermosphere.

The second interesting feature is that the SET HASDM density data display a common range of variability despite the level of daily averaged geomagnetic activity. During higher levels of daily averaged geomagnetic activity, the density mean and median values increase at all altitude levels. However, the relative range of variability is consistent from one daily average of Ap to the next. The reason for this is likely to be that the underlying pre-storm density of the thermosphere is determined by the solar EUV and FUV irradiances that create a thermal foundation of the upper atmosphere. The daily averaged geomagnetic activity is a perturbation upon that foundation. The global range of variation occurs because of geomagnetic activity perturbations separate from the underlying thermal inertia in the atmosphere. The mean and median density increases are related to geomagnetic storms.

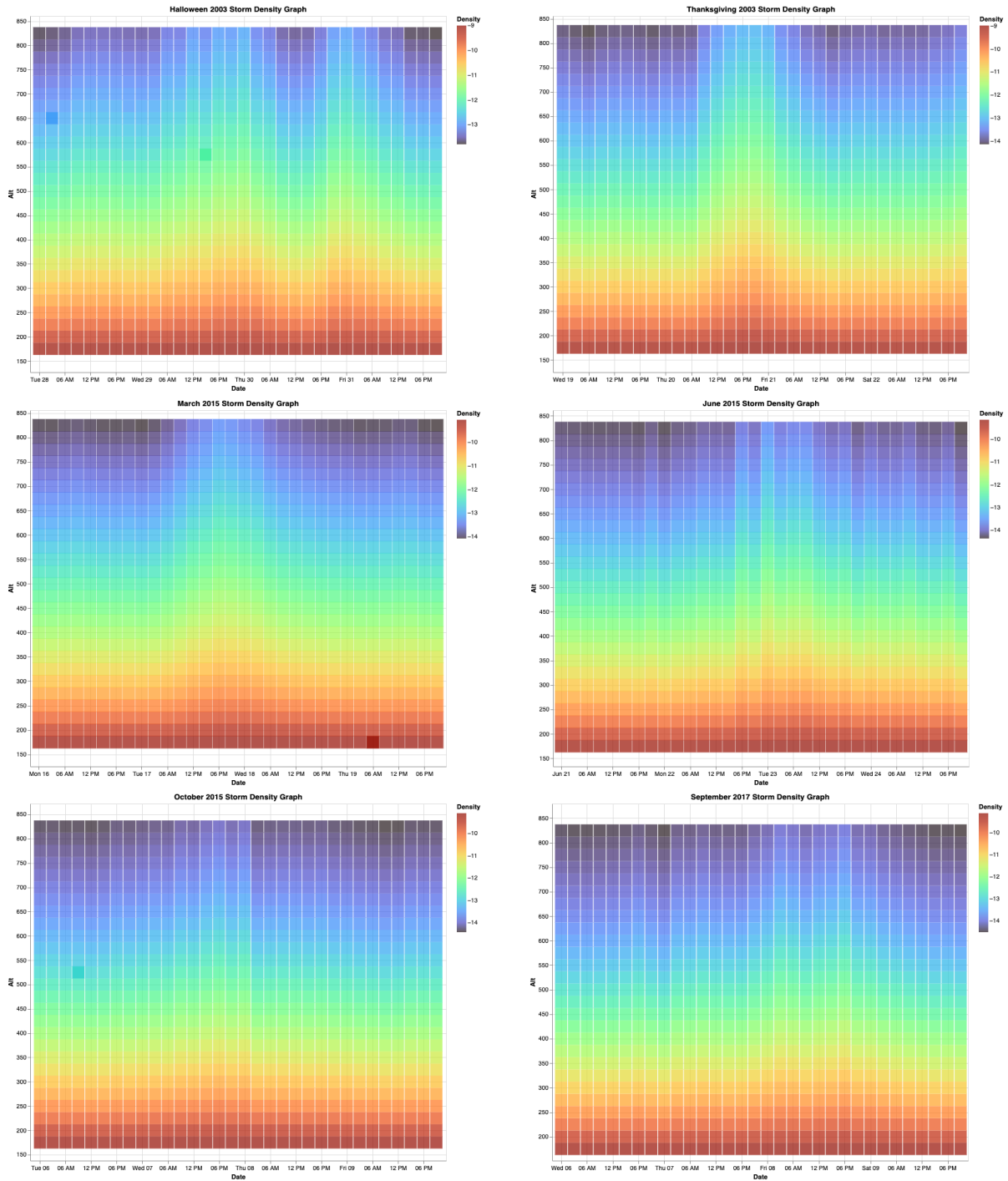

Fig. 10a-f. Individual storm densities by altitude and time.

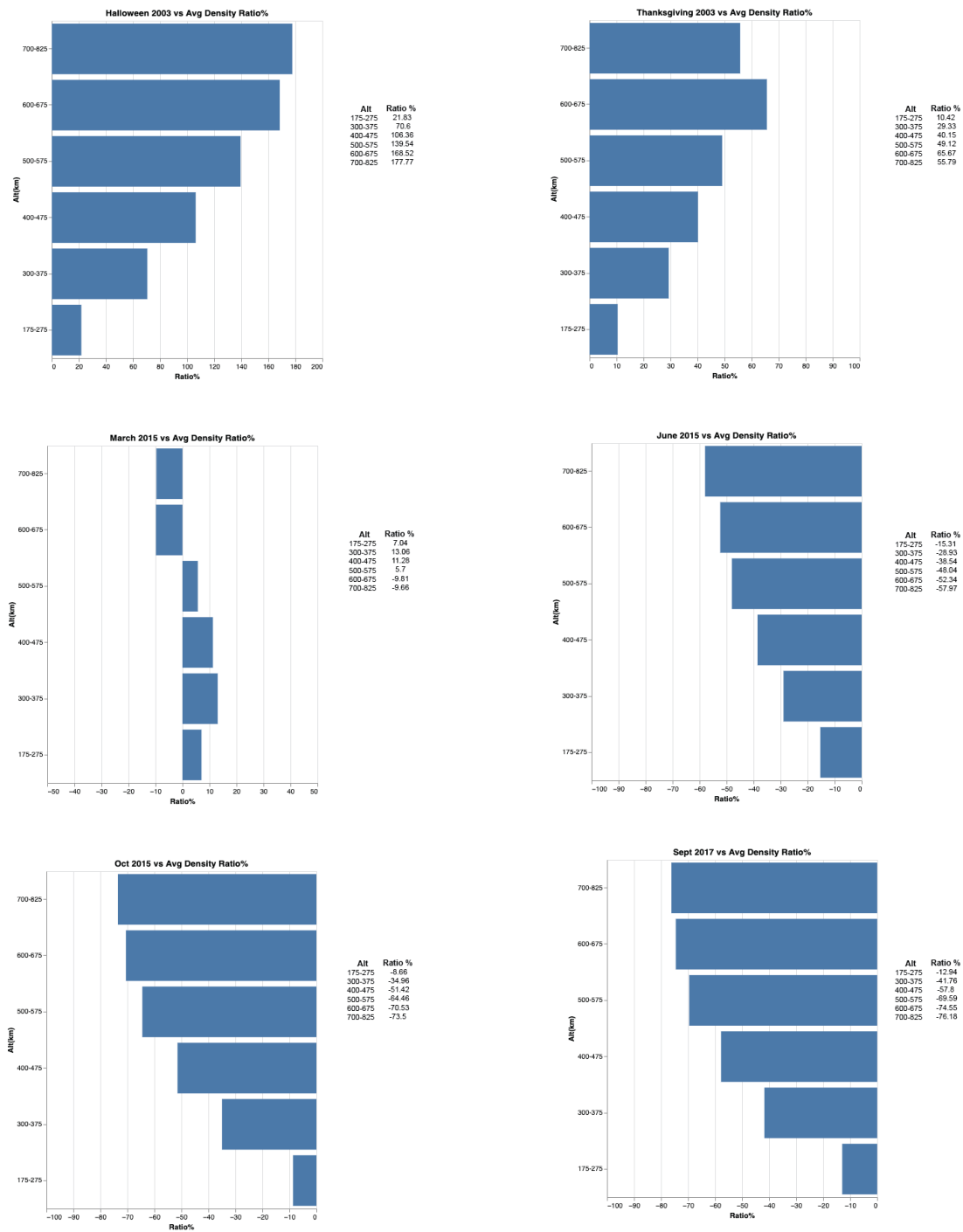

Fig. 11a-f. Individual storm density ratios (percentage) to average storm values by altitude.

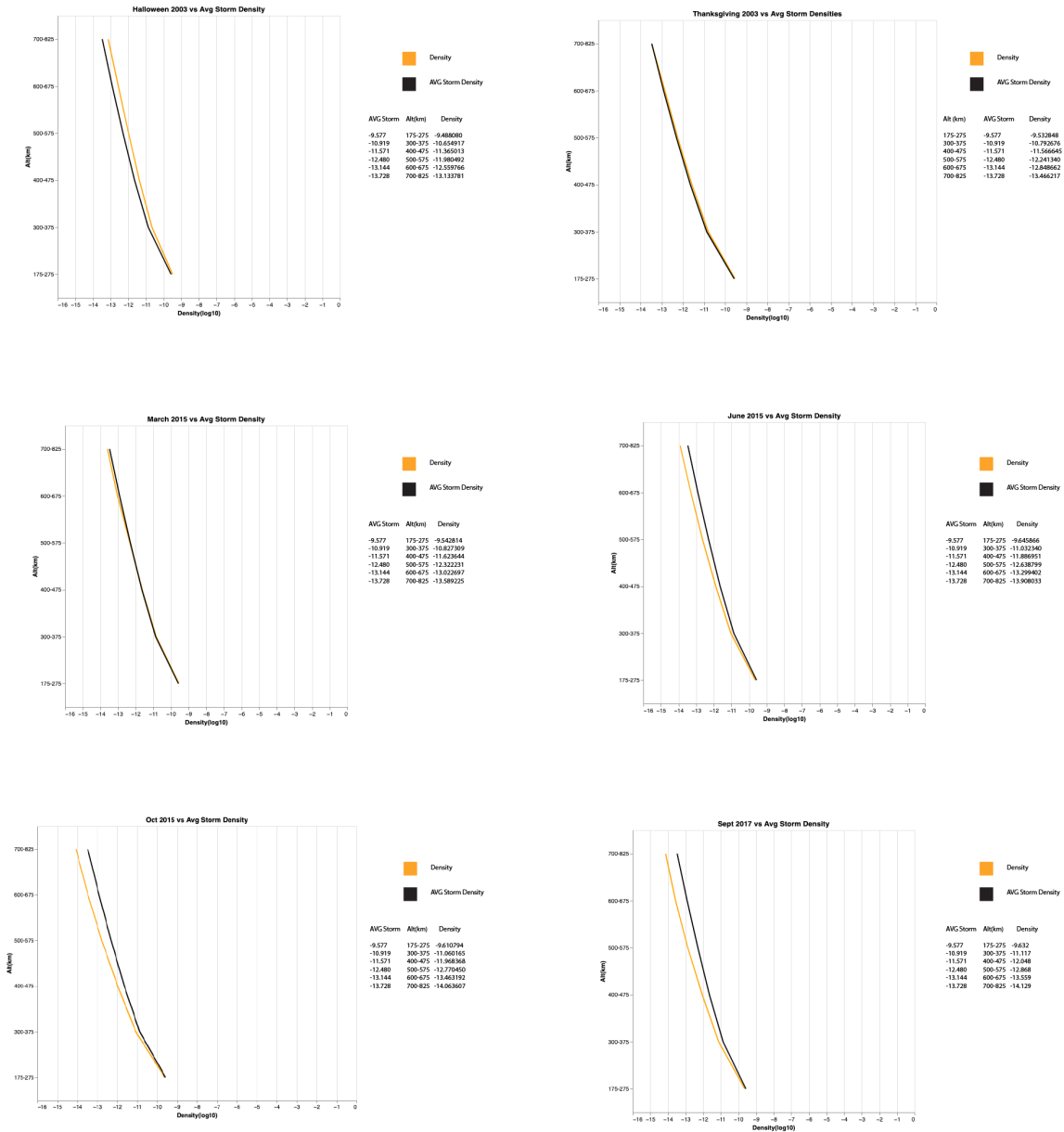

Fig. 12a-f. Individual storm density ratios (gold) to average storm values (black) by altitude.

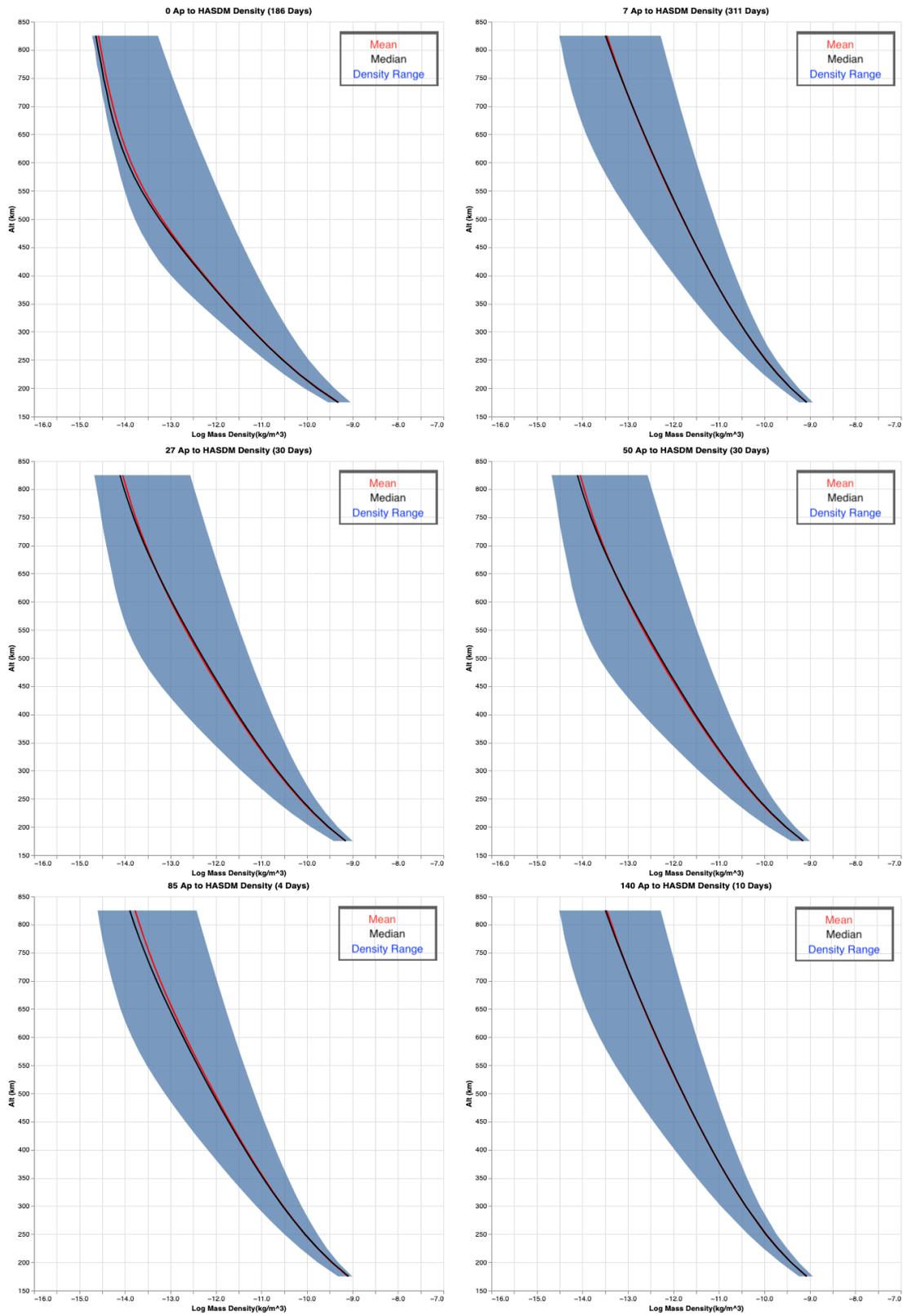

Fig. 13a-f. Neutral density variability for daily average Ap 0, 7, 27, 50, 85, and >140.

## 5. REFERENCES

[1] *National Orbital Debris Research and Development Plan* (2021), A Report by the Orbital Debris Research and Development Interagency Working Group Subcommittee on Space Weather, Security, and Hazards Committee on Homeland and National Security of the National Science and Technology Council.

[2] *ISO 24113:2011 Space Systems – Space debris mitigation requirements*, International Standards Organization, Geneva, 2011.

[3] Hejduk, M. D. and D. E. Snow (2018). The effect of neutral density estimation errors on satellite conjunction serious event rates. *Space Weather*, 16, 849–869. https://doi.org/10.1029/2017SW001720.

[4] Storz, M. F., B. R. Bowman, M. J. I. Branson, S. J., Casali, and W. K. Tobiska (2005). High accuracy satellite drag model (HASDM). *Advances in Space Research*, *36*(12), 2497-2505.

[5] Tobiska, W.K., B.R. Bowman, D. Bouwer, A. Cruz, K. Wahl, M. Pilinski, P.M. Mehta, and R.J. Licata (2021), The SET HASDM database, *Space Weather*, *19*, e2020SW002682, https://doi. org/10.1029/2020SW002682.

[6] Tobiska, K., B. Bowman, M. D. Pilinski, P. Mehta, and R. Licata (2021), The Machine Learning Enabled Thermosphere Advanced by the High Accuracy Satellite Drag Model (META-HASDM), AMOS conference, Maui Hawaii, 2021.

[7] Space Weather Operations, Research, and Mitigation Working Group; Space Weather, Security, and Hazards Subcommittee; Committee on Homeland and National Security of the National Science and Technology Council, The White House, *National Space Weather Strategy and Action Plan*, March 2019.

[8] Licata, R.J., P.M. Mehta, W.K. Tobiska, B.R. Bowman, and M.D. Pilinski (2021a), Qualitative and quantitative assessment of the SET HASDM database, *Space Weather*, *19*, e2021SW002798, https://doi. org/10.1029/2021SW002798.

[9] Licata, R.J., P.M. Mehta, W.K. Tobiska, and S.V. Huzurbazar (2021b), Machine- learned HASDM thermospheric mass density model with uncertainty quantification, *Space Weather*, *20*, e2021SW002915, https://doi. org/10.1029/2021SW002915.

[10] Banks, P.M. and G. Kockarts (1973), *Aeronomy*, Academic Press, New York.

[11] Bowman, B.R., W.K. Tobiska, F.A. Marcos, C.Y. Huang, C.S. Lin, W.J. Burke, A New Empirical Thermospheric Density Model JB2008 Using New Solar and Geomagnetic Indices (2008), *AIAA/AAS Astrodynamics Specialist Conference*, AIAA 2008-6438.

[12] Tobiska, W.K., S.D. Bouwer, and B.R. Bowman (2008), The development of new solar indices for use in thermospheric density modeling, *JASTP*, 70, 803-819.

[13] Licata, R.J., P.M. Mehta, D.R. Weimer, D.P. Drob, W.K. Tobiska, and J. Yoshii (2022), Science Through Machine Learning: A Case Study for Post-storm Thermospheric Cooling, arXiv:2206.05824 [physics.space-ph].